\documentclass[12pt,a4paper]{article}

\usepackage{arxiv}
\usepackage{xfrac}
\usepackage{graphicx}
\usepackage{amsmath,amssymb,amsfonts}
\usepackage{subcaption}
\usepackage{makecell}
\usepackage{xcolor}
\usepackage{textcomp}
\usepackage[numbers, sort&compress]{natbib}

\onecolumn

\makeatletter
\def\ps@pprintTitle{%
  \fancyhf{}%
  \renewcommand{\headrulewidth}{0pt}%
  \renewcommand{\footrulewidth}{0pt}%
  \fancyhead[C]{}%
  \fancyhead[R]{}%
  \fancyhead[L]{}%
  \fancyfoot[C]{\thepage}%
  \pagestyle{plain}%
}
\makeatother

\AtBeginDocument{%
  \onecolumn
  \pagestyle{plain}%
}

\usepackage[utf8]{inputenc}
\usepackage[T1]{fontenc}
\usepackage{hyperref}
\usepackage{url}

\title{Investigation of Soft Magnetic Composites in Radial‑Flux Wound Field Synchronous Machines for Automotive Propulsion}

\author{
  Andreas Carlsson \textsuperscript{1}, Christian Sandström \textsuperscript{1}, Viktor Josefsson \textsuperscript{1}, Lisa Kjellén \textsuperscript{2}, \\
  \textbf{Taha El Hajji \textsuperscript{3}, Marcus Lenberg \textsuperscript{2}} \\
  \textsuperscript{1} Electric Drive Unit, Polestar Performance AB, Gothenburg, Sweden \\
  \textsuperscript{2} Electro Magnetic Materials, Höganäs AB, Höganäs, Sweden \\
  \textsuperscript{3} Alvier Mechatronics AB, Helsingborg, Sweden \\
}

\begin{document}
\let\preprintfooter\relax
\maketitle

\begin{abstract}
This paper investigates the application of soft magnetic composites (SMCs) in the stators of wound field synchronous machines for automotive traction. While SMCs are traditionally employed in axial flux topologies, this study examines their use in radial-flux electrically excited synchronous machines (EESMs). Multiple SMC materials and lamination thicknesses are evaluated, with the optimal configuration combining a SMC material in the stator and 0.35 mm NO35 laminated steel in the rotor. This combination delivers improved torque and efficiency compared to conventional designs. When integrated into a full electric drive unit (EDU), this motor achieves 89.7\% efficiency over the WLTP drive cycle, representing a 1.4 percentage point improvement over a reference permanent magnet synchronous machine-based EDU. The proposed solution eliminates rare-earth materials, reduces cost through thicker laminations, and offers environmental benefits through SMC utilization. This novel material combination, previously unexplored for radial EESMs, presents a promising direction for affordable, high-efficiency, rare-earth-free automotive traction machines.
\end{abstract}


\section{Introduction}

The automotive industry is undergoing a significant transformation as internal combustion engines are phased out in favor of electric powertrains—a critical step toward reducing carbon emissions that remains a high priority for policymakers, researchers, and industry stakeholders. Automotive OEMs have invested heavily in high-performance electric propulsion over recent decades. Since the high-voltage battery represents the most cost-intensive component in electric vehicles, improving powertrain efficiency enables battery downsizing, delivering substantial cost savings for the complete vehicle.

Historically, permanent magnet synchronous machines (PMSMs) have been the machine type of choice for most OEMs due to their strong performance characteristics and high efficiency. These machines are also relatively straightforward to design, control, and manufacture, making them an attractive option for automotive applications. One of the principal disadvantages of this machine type, however, is its dependence on rare-earth magnets, which are typically required for their high remanence flux and coercive force that ensure reliability across a wide operating temperature range. There are significant environmental concerns associated with rare-earth magnet production, and additionally, the pricing of these components remains both high and unreliable due to market volatility and supply chain constraints. These factors have created a pressing need for rare-earth-free motor solutions that can deliver comparable performance while reducing environmental impact and material costs.

One alternative being actively explored by engineers is the utilization of wound field synchronous machines (WFSMs) \cite{1,2} as replacements for conventional PMSMs \cite{3}. The primary advantage of this machine type lies in its use of an electromagnet in the rotor instead of permanent magnets, which eliminates the need for rare-earth elements. WFSMs can typically achieve increased power output compared to conventional PMSMs and maintain higher power levels even when operating in field weakening conditions. They also exhibit improved efficiency at low torque operating points and incur no electromagnetic losses under no-load conditions. The ability to actively control the rotor field enables these performance improvements relative to PMSMs. However, this technology presents several disadvantages as well, with the main challenge being thermal management due to the introduction of losses in the rotor, which may necessitate active cooling systems. Another drawback associated with the rotor is that these additional losses make this motor type less efficient at low speeds and medium to high torque levels compared to PMSMs. WFSMs \cite{4} typically operate with higher stator current magnitudes at these operating points because the rotor magnetic field is generally not as strong as that of a PMSM at medium torque levels when the field current is not increased to its maximum value. This characteristic means that losses generated by the power electronics can also be higher compared to those in PMSM-based systems.

The introduction of field current as a new control variable presents additional challenges for control system design, as conventional maximum torque per ampere (MTPA) algorithms commonly used for automotive electrical machines cannot be applied in their standard form, as they would lead the motor to operate at peak field current across nearly all operating points. Consequently, alternative algorithms such as maximum torque per loss (MTPL) or maximum torque per copper loss (MTPCuL) must be implemented. There is also the challenge of transferring electrical energy to the rotor to energize the field winding, for which several solutions have been proposed. Historically, mechanical brushes have been employed, offering the advantages of simplicity and accurate control, but introducing mechanical friction losses. Researchers and engineers have recently proposed alternative methods for power transfer to the rotor, including wireless power transfer using rotating transformers or rotating supercapacitors. The rotating transformer approach has been studied extensively and is currently employed as a viable alternative that improves system reliability while eliminating mechanical friction losses. This wireless solution does, however, introduce complexity and additional components, and may require advanced current estimation.

The magnetic core of an electrical machine is fundamental to its torque production and efficiency, with material selection presenting a critical choice for designers \cite{5,6,7}. Traditional Soft Magnetic Steels (SMS), manufactured as thin, insulated sheets stacked to form stator and rotor cores, have long served as the industry standard due to their well-understood characteristics and established manufacturing processes. These laminated steels offer predictable magnetic behavior and low eddy current losses when properly designed, making them a reliable choice for conventional machine topologies. In this investigation, we specifically examine 0.35 mm thickness laminations as a cost-effective material option that will be shown to be the best option for the rotor core. Soft Magnetic Composites (SMCs) \cite{8}, produced through powder metallurgy techniques, offer a compelling alternative for the stator core allowing different permeability and saturation levels. Beyond their technical advantages, SMCs \cite{9} represent a potentially more ecological material solution, as their manufacturing process can reduce material waste and they offer greater design flexibility that may ultimately contribute to more sustainable motor production. SMC, traditionally used for axial flux motors, is used in this context for radial flux motors.

When comparing conventional PMSMs and WFSMs, several key considerations emerge that engineers must evaluate when determining the most suitable motor type for a given application. The vehicle type, powertrain layout, and intended operating profile are crucial factors in selecting the appropriate machine technology. For instance, vehicles operating predominantly at higher torque levels and lower speeds, such as heavy-duty machinery and utility vehicles, may benefit from PMSM motors, while vehicles that primarily operate at high speed and low torque, such as those engaged in highway driving, may find WFSMs more advantageous. Since WFSMs exhibit less favorable efficiency characteristics at medium to high torque levels, partly due to operation at higher stator currents, maximizing the number of equivalent turns in the stator winding is beneficial as it allows the motor to achieve the same magnetomotive force at lower current levels. As WFSMs typically demonstrate higher power output in field weakening compared to similarly designed PMSMs, employing a higher number of stator turns relative to the PMSM design can be a viable optimization strategy. In the case of the reference machine presented in Section 3, the design incorporates the maximum number of turns possible while still meeting peak power requirements, a deliberate design choice made to maximize efficiency at medium and high torque levels during low-speed operation. This paper investigates how WFSMs utilizing both SMC and cost-effective 0.35 mm laminated steels can achieve high efficiency through considerable reduction in stator current, thereby addressing the dual imperatives of cost reduction and sustainable motor design for next-generation automotive applications.

The paper is organized as follows. Section 2 presents a background of both soft magnetic steels (SMS) and composites (SMC). Section 3 presents the reference machine used for benchmarking. Section 4 and Section 5 detail the driving cycle analysis and the design methodology, respectively. Finally, Section 6 presents the optimization and benchmark results.

\section{Soft Magnetic Steel and Composites}
The stator and rotor cores of electrical machines have traditionally been manufactured from laminated Soft Magnetic Steels (SMS). In production, these laminations are typically punched from continuous rolls of steel sheet, requiring material properties suitable for both components. Utilizing the same steel grade for stator and rotor reduces costs, manufacturing complexity, and material waste. Individual laminations—typically 0.2 to 0.35 mm to minimize eddy current losses—are coated with an insulating layer. These laminations are subsequently stacked and consolidated into a complete core using methods such as welding, bonding, or interlocking features. The insulating coating occupies a portion of the stack length, meaning that not all of the active length consists of magnetic material. The ratio of actual magnetic material to total stack length is termed the stacking factor, typically ranging from 94\% to 96\%, and is specified in steel manufacturer datasheets.

A variety of soft magnetic steel grades are employed in automotive traction machines. The steel must be non-oriented to ensure isotropic magnetic properties, with common alloy systems including iron-silicon (Fe-Si), iron-nickel (Fe-Ni), and iron-cobalt (Fe-Co). Each alloy family offers distinct magnetic characteristics: silicon steels are widely used for their low core losses and cost-effectiveness; cobalt-iron alloys achieve significantly higher saturation flux density, enabling greater torque density at the expense of substantially higher material cost; nickel-iron alloys provide extremely high permeability but lower saturation. In addition to alloy composition, lamination thickness critically influences performance—thinner laminations reduce eddy current losses but slightly degrade stacking factor and increase manufacturing cost. The cost disparity between steel types can be substantial, with cobalt-iron alloys costing up to 100 times more than conventional silicon steel, and thinner gauge materials commanding premium pricing. For automotive applications requiring cost-effective solutions, 0.35 mm silicon steel laminations represent a widely adopted baseline that balances electromagnetic performance with manufacturability and cost.

Soft Magnetic Composites (SMCs) represent a fundamentally different class of soft magnetic materials manufactured through powder metallurgy. These materials consist of high-purity iron particles, each coated with a thin inorganic insulating layer that electrically isolates individual particles to minimize eddy current losses. The coated powder is compacted under high pressure—typically 400 to 1100 MPa—using precision tooling to produce near-net-shape components. Following compaction, a controlled heat treatment relieves internal stresses introduced during pressing, restoring magnetic softness and optimizing permeability. The resulting material exhibits isotropic magnetic and thermal properties, a key enabler for three-dimensional flux paths impossible to achieve with laminated steels. Commercial SMC grades include Somaloy 700HR 3P, Somaloy 700HR 5P, and X-Somaloy 700 7P from Höganäs, among others. The density of the finished component directly governs its magnetic performance: higher compacted density yields a steeper B-H curve, higher permeability, and greater magnetic flux density for a given magnetizing field. Saturation flux density exhibits a linear relationship with density, while improved inter-particle contact at higher densities also reduces core losses. However, achieving higher density demands exponentially greater pressing force, imposing practical constraints based on available press capacity and tooling strength.

Unlike laminated steels, SMC components are manufactured to final shape in a single pressing operation, producing a homogeneous magnetic core with no stacking factor penalty—the entire volume consists of active magnetic material. However, this manufacturing approach introduces unique considerations for automotive applications. The pressing force required scales with component cross-sectional area, meaning that larger stator diameters demand proportionally higher press forces, potentially exceeding the capacity of available manufacturing equipment. Furthermore, the tooling investment for SMC compaction is substantial, requiring high production volumes to achieve economic viability for automotive OEMs. The pressed components are brittle in their green state and require careful handling prior to heat treatment, which imparts final mechanical integrity. The analysis focuses on three SMC materials, identified here as SMC A, SMC B, and SMC C to preserve confidentiality.

\section{Reference Motor}
To establish a baseline for investigating the application of soft magnetic composites in wound field synchronous machines, two reference machines are considered in this study: a conventional permanent magnet synchronous machine representing established automotive technology, and a wound field synchronous machine serving as the platform for material investigation.

\begin{figure}[htbp]
\centerline{\includegraphics[width=0.5\columnwidth]{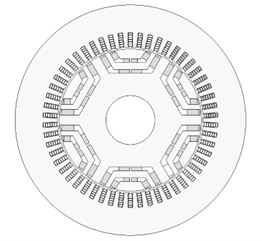}}
\caption{PMSM motor 2U topology.}
\label{fig:1}
\end{figure}

The reference PMSM is a state-of-the-art traction motor designed for automotive applications, delivering 270 kW of peak power, as shown in Fig.~\ref{fig:1}. This machine utilizes a typical automotive-grade sintered NdFeB magnet with a remanence flux density of 1.3 T, representing conventional rare-earth based permanent magnet technology. The motor is operated by a 530 Arms inverter, consistent with high-performance electric vehicle powertrains. Comprehensive specifications of this reference machine are presented in Table~\ref{tab:1}. This PMSM serves as the performance benchmark against which the WFSM variants will be evaluated, providing a reference point for assessing the trade-offs associated with eliminating rare-earth materials.

\begin{figure}[htbp]
\centerline{\includegraphics[width=0.5\columnwidth]{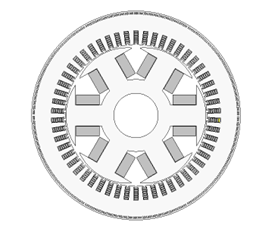}}
\caption{WFSM outline and topology.}
\label{fig:2}
\end{figure}

The wound field synchronous machine investigated in this study is a radial-flux design rated for 210 kW peak power, as illustrated in Fig.~\ref{fig:2}. This machine is powered by a 440 Arms silicon carbide (SiC) inverter, offering improved switching characteristics and efficiency compared to conventional IGBT-based drives. The excitation system employs a contactless rotating transformer capable of transferring up to 8.2 kW to the rotor field winding, eliminating the need for mechanical brushes and their associated friction losses while enabling precise control of the rotor magnetic field. The stator and rotor cores of this WFSM are currently manufactured from a commercially available 0.25 mm non-oriented silicon steel, representing conventional laminated steel technology. This machine therefore provides a suitable platform for investigating the replacement of laminated steels in the stator with soft magnetic composites and the replacement of laminated steels in the rotor with thicker laminations. Detailed specifications of this WFSM are also provided in Table~\ref{tab:1}.

\begin{table}[htbp]
\caption{PMSM and WFSM motor parameters}
\label{tab:1}
\begin{center}
\begin{tabular}{|l|c|c|c|}
\hline
\textbf{Parameter} & \textbf{WFSM (M0)} & \textbf{PMSM} & \textbf{Unit} \\
\hline
Peak Torque & 610 & 450 & N.m \\
Peak Power & 210 & 270 & kW \\
Rated Voltage & 625 & 625 & V \\
Maximum Voltage & 900 & 900 & V \\
Maximum Current & 400 & 550 & Arms \\
Field Current & 32 & N/A & A \\
Stator Outer Diameter & 237 & 237 & mm \\
Active Length & 133 & 128 & mm \\
Lamination & NO25 & NO25 & N/A \\
\hline
\end{tabular}
\end{center}
\end{table}

\section{Drive Cycle Analysis}
To ensure the investigation reflects realistic automotive operating conditions, the final performance comparison is conducted at the electric drive unit (EDU) level for a small passenger vehicle. The vehicle platform employs an 800 V battery architecture and features rear-wheel drive configuration. A key distinction between the two machine types lies in their transmission systems: the WFSMs utilize a single-stage planetary gear set with a ratio of 7.0, while the PMSM employs a dual-stage transmission with a ratio of 9.5. This difference was necessitated by the PMSM's lower torque capability and was specifically selected to minimize overall powertrain losses during highway driving conditions.

\begin{figure}[htbp]
\centerline{\includegraphics[width=0.5\columnwidth]{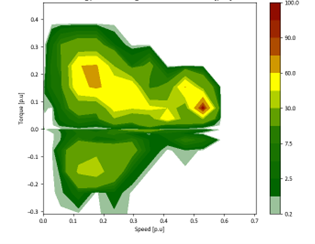}}
\caption{WLTP EDU energy consumption map over the full cycle. Color bar represents the normalized energy consumption per operating point in the torque/speed map.}
\label{fig:3}
\end{figure}

The energy consumption was analyzed using a vehicle simulation tool under the Worldwide Harmonized Light Vehicles Test Procedure (WLTP) cycle. From this analysis, two critical operating points were identified as representing the regions of highest energy dissipation during typical vehicle operation, as illustrated in Fig.~\ref{fig:3}. These two hotspots serve as the primary optimization targets for the subsequent material investigation. The first operating point, designated OP1, represents an acceleration event characteristic of urban driving conditions. The second operating point, OP2, corresponds to constant-speed highway cruising. In addition to these efficiency-focused operating points, the design and optimization process incorporates fundamental machine requirements including peak torque capability, peak power delivery, and DC voltage constraints. Weight coefficients are assigned to each constraint and optimization objective to reflect their relative importance in the final machine design, with these coefficients summarized in Table~\ref{tab:2}. This multi-objective approach ensures that material improvements targeting cycle efficiency do not compromise the machine's ability to meet its fundamental performance requirements.

\begin{table}[htbp]
\caption{Optimization objectives/constraints}
\label{tab:2}
\begin{center}
\begin{tabular}{|l|c|c|}
\hline
\textbf{Characteristics} & \textbf{Objective / Constraint} & \textbf{Weight (1-10)} \\
\hline
OP1 Torque & $>$145 Nm & 6 \\
OP1 Efficiency & Maximize & 8 \\
OP2 Torque & $>$40 Nm & 6 \\
OP2 Efficiency & Maximize & 7 \\
Peak Torque & $>$550 Nm & 10 \\
Peak Power & $>$200 kW & 8 \\
DC Voltage & $<$625 V & 8 \\
\hline
\end{tabular}
\end{center}
\end{table}

\section{Design Methodology}
This investigation considers six distinct machine variants to comprehensively evaluate the impact of soft magnetic materials on WFSM performance. For the WFSM platform, the three SMC materials under investigation—designated SMC A, SMC B, and SMC C—are applied to the stator core, while two different soft magnetic steel thicknesses are considered for the rotor core: NO25 (0.25 mm) and NO35 (0.35 mm). The NO35 lamination is a commercially available 0.35 mm silicon steel which exhibits improved inductive properties compared to the reference machine steel, albeit with higher core losses. This material also offers a cost advantage over the thinner NO25 laminations, making it an economically attractive alternative for automotive applications where material cost is a primary concern. Table~\ref{tab:3} provides a comprehensive summary of all reference and benchmarked motor configurations considered in this study.

The WFSM stator design required several modifications to accommodate the characteristics of SMC materials. First, the tooth tips were eliminated, resulting in open slots, a necessary change imposed by manufacturing limitations of SMC compacting tools which cannot readily produce undercut features. Second, because SMC materials exhibit different permeability characteristics and saturation flux densities compared to conventional NO25 silicon steel, the slot dimensions—including width, depth, and yoke width—were adjusted accordingly to optimize magnetic performance. The stator outer diameter, active length, and winding layout were deliberately maintained unchanged from the reference design to isolate the effect of material substitution. The rotor geometry was also preserved without modification to focus the investigation on stator material influence.

Design optimization was conducted using JMAG, a finite element analysis (FEA) solver, employing its built-in multi-objective genetic algorithm. The optimization models were configured to balance computational efficiency with acceptable accuracy in loss calculations. The machine performance evaluation incorporated iron losses, DC copper losses, and AC copper losses to capture the full loss spectrum under operating conditions.

Following completion of the optimization runs, the resulting Pareto fronts were analyzed, and the most promising design candidates were selected from each material variant. These selected designs were subsequently re-evaluated using a more sophisticated 2D modeling approach that considers the complete electric drive unit, including transmission losses and power electronics losses, through an internal high-accuracy modeling tool. This comprehensive EDU-level analysis ensures that system-level interactions are properly captured. It should be noted that induced PWM losses in the motor were not explicitly modeled; instead, an industry-standard correction factor for losses was applied consistently to both motor types. The control strategy employed for all WFSM variants was Maximum Torque Per Loss (MTPL), which optimally balances the trade-off between stator and field winding losses. The final candidate designs were then compared against the conventional PMSM-based EDU using total WLTP drive cycle efficiency as the primary performance metric.

\begin{table}[htbp]
\caption{Reference and benchmarked motors}
\label{tab:3}
\begin{center}
\begin{tabular}{|l|c|c|}
\hline
\textbf{Motors} & \textbf{Stator Core} & \textbf{Rotor Core} \\
\hline
PMSM & NO25 & NO25 \\
WFSM M0 & NO25 & NO25 \\
WFSM M1 & SMC A & NO25 \\
WFSM M2 & SMC B & NO25 \\
WFSM M3 & SMC C & NO25 \\
WFSM M4 & SMC A & NO35 \\
WFSM M5 & SMC B & NO35 \\
WFSM M6 & SMC C & NO35 \\
\hline
\end{tabular}
\end{center}
\end{table}

\section{Results}
Following the optimization process described in Section 5, multiple feasible motor designs were obtained for each of the six variants (M1 to M6), distributed along the Pareto front representing the trade-off between competing objectives i.e., efficiency at OP1 and efficiency at OP2. From each Pareto front, the most promising candidate was selected using an internal evaluation framework. While the exact selection criteria are confidential, the general approach can be described as follows: all viable candidates were extracted from the optimization results and subjected to post-processing to estimate their total business value (TBV). This comprehensive metric incorporates material costs \cite{10}, production costs, the monetary value of efficiency improvements, and the value of performance enhancements. The candidate yielding the highest TBV was selected for further investigation from each material variant. The final selected designs are presented in Table~\ref{tab:4}, which summarizes the steel configurations, performance characteristics, and efficiency metrics for each candidate.

\begin{table}[htbp]
\caption{Benchmark of reference motor and optimized motors}
\label{tab:4}
\begin{center}
\begin{tabular}{|l|c|c|c|c|}
\hline
\textbf{Motors} & \textbf{Peak Torque} & \textbf{Peak Power} & \textbf{OP1 Eff.} & \textbf{OP2 Eff.} \\
\hline
PMSM & 450 N.m & 270 kW & 93\% & 93.3\% \\
WFSM M0 & 610 N.m & 210 kW & 90.7\% & 94.8\% \\
WFSM M1 & 554 N.m & 207 kW & 89.7\% & 93.6\% \\
WFSM M2 & 607 N.m & 210 kW & 90.5\% & 94.6\% \\
WFSM M3 & 610 N.m & 210 kW & 91\% & 94.9\% \\
WFSM M4 & 572 N.m & 210 kW & 89.9\% & 93.6\% \\
WFSM M5 & 635 N.m & 215 kW & 91.3\% & 94.6\% \\
WFSM M6 & 640 N.m & 215 kW & 91.5\% & 94.8\% \\
\hline
\end{tabular}
\end{center}
\end{table}

Among all optimized variants, the WFSM designated M6 emerged as the top performer in terms of efficiency. This configuration utilizes SMC C in the stator core combined with NO35 (0.35 mm) laminated steel in the rotor. For reference, the baseline WFSM (designated M0) using conventional NO25 steel in both stator and rotor is included in Table~\ref{tab:4} for comparison, though this design was not carried forward into the subsequent evaluation phase.

The M6 motor was subsequently integrated into a complete electric drive unit model for comprehensive system-level analysis. The energy consumption over the full WLTP drive cycle was evaluated and compared against the reference PMSM-based EDU. Table~\ref{tab:5} presents the results of this comparative analysis. The EDU utilizing the M6 WFSM achieved a WLTP cycle efficiency of 89.7\%, representing a significant improvement over the reference PMSM-based EDU which achieved 88.3\% efficiency. This corresponds to an absolute efficiency increase of 1.4 percentage points at the EDU level. Additionally, the efficiency at two constant-speed operating points—70 km/h and 130 km/h—was evaluated, with the M6 motor demonstrating substantially higher efficiency at both points compared to the reference PMSM. The primary efficiency improvement stems from reduced current in the WFSM, which decreases the Joule losses that dominate loss generation in this machine type.

\begin{table}[htbp]
\caption{EDU efficiency over WLTP drive cycle: reference PMSM vs. WFSM-M6}
\label{tab:5}
\begin{center}
\begin{tabular}{|l|c|c|}
\hline
\textbf{Characteristic} & \textbf{PMSM} & \textbf{WFSM M6} \\
\hline
\textbf{WLTP Efficiency} & \textbf{88.3\%} & \textbf{89.7\%} \\
70 km/h Efficiency & 88.5\% & 91.7\% \\
130 km/h Efficiency & 89.9\% & 92.5\% \\
\hline
\end{tabular}
\end{center}
\end{table}

These results yield several important conclusions. First, the combination of SMC material in the stator and thicker NO35 laminations in the rotor delivers measurable efficiency gains despite the lower material cost of the thicker laminations. The 1.4 percentage point improvement in WLTP cycle efficiency at the EDU level demonstrates that cost-effective material choices need not compromise performance. Second, the M6 motor achieves these improvements while completely eliminating rare-earth materials from the machine construction, positioning it as a viable rare-earth-free alternative for automotive traction applications. Third, from a cost perspective, the elimination of expensive permanent magnets combined with the use of lower-cost NO35 rotor laminations offers substantial cost reduction potential. Finally, from a sustainability standpoint, the utilization of SMC materials—which can be manufactured with reduced environmental impact compared to traditional steel processing—combined with the avoidance of rare-earth mining and processing, positions this motor configuration as an environmentally responsible, low-cost, rare-earth-free, and high-efficiency option for next-generation electric vehicles.

\section{Conclusion}
This paper has investigated the utilization of soft magnetic composite stators in wound field synchronous machines for automotive traction applications, demonstrating the potential of both commercially available SMC materials and emerging grades. While SMCs have traditionally been employed in axial flux topologies, this study explores their application in radial-flux electrically excited synchronous machines, representing a novel direction for this material class.

Multi-objective optimizations were conducted to design WFSMs suitable for a typical passenger vehicle, evaluating multiple SMC materials in the stator combined with different lamination thicknesses in the rotor. Among the configurations investigated, the combination of SMC C in the stator with 0.35 mm NO35 laminated steel in the rotor emerged as the most promising. A key advantage of this approach is the ability to independently select rotor steel for optimal properties without stator loss constraints—NO35 offers improved inductive characteristics and lower cost compared to thinner laminations.

The selected WFSM design was integrated into a full electric drive unit model and benchmarked against a reference permanent magnet synchronous machine-based EDU. Over the WLTP drive cycle, the SMC-based WFSM achieved 89.7\% efficiency, representing a 1.4 percentage point improvement over the PMSM reference. Superior efficiency was also demonstrated at constant-speed operating points of 70 km/h and 130 km/h. These efficiency gains are primarily driven by current reduction in the WFSM, minimizing the dominant Joule losses.

The proposed solution delivers multiple simultaneous benefits: complete elimination of rare-earth materials, cost reduction through thicker NO35 rotor laminations, high efficiency across the drive cycle, and environmental advantages through SMC utilization. This novel material combination—SMC stators with 0.35 mm laminated rotors in radial EESMs—has not been previously explored and represents a promising direction for affordable, high-efficiency, rare-earth-free traction machines for next-generation electric vehicles.

\section*{Acknowledgment}
The authors would like to thank Polestar Performance AB for providing the facilities, software and computational hardware making this research possible.

\bibliographystyle{unsrtnat}
\bibliography{references}

\end{document}